\documentclass[aps,prl,twocolumn,groupedaddress]{revtex4}
\usepackage{graphicx}
\usepackage{amsmath}
\usepackage{natbib}

\newcommand{\fc}{{f_c}}           
\newcommand{\fd}{{f^*}}           
\newcommand{\lss}{{\ell_{\rm s}}}   
\newcommand{\lds}{{\ell_{\rm d}}}   
\newcommand{\eps}{{\varepsilon_{\rm b}}}   
\newcommand{\El}{{E_{\ell}}}        
\newcommand{\Eini}{{\varepsilon_{\ell}}}  
\newcommand{\cS}{{\cal S}}        
\newcommand{\nr}{{m}}               
\newcommand{\bss}{b_{\rm s}}   
\newcommand{\bds}{b_{\rm d}}   
\newcommand{\Zh}{\hat{Z}}       
\newcommand{\Tr}{\tau}   
\newcommand{\cP}{{\cal P}}      

\begin{document}

\title{Dynamics of force-induced DNA slippage}

\author{Richard A. Neher} 
\email{Richard.Neher@physik.lmu.de}

\author{Ulrich Gerland}
\email{Ulrich.Gerland@physik.lmu.de}

\affiliation{Department of Physics and CENS, LMU M\"unchen, 
Theresienstrasse 37, 80333 M\"unchen, Germany}

\date{\today}

\begin{abstract} 
We study the basepairing dynamics of DNA with repetitive sequences where 
local strand slippage can create, annihilate, and move bulge loops. 
Using an explicit theoretical model, we find a rich dynamical behavior 
as a function of an applied shear force $f$: reptation-like dynamics at 
$f=\fc$ with a rupture time $\tau$ scaling as $N^3$ with its length $N$,  
drift-diffusion dynamics for $\fc<f<\fd$, and a {\em dynamical} transition 
to an unraveling mode of strand separation at $f=\fd$. 
We predict a viscoelastic behavior for periodic DNA with 
time and force scales that can be {\em programmed} into its sequence. 
\end{abstract}

\pacs{???}
\maketitle

\begin{figure}[b]
\includegraphics[width=8cm]{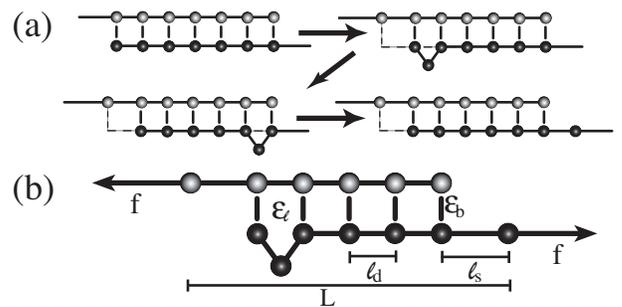}
\vspace*{-0.4cm}
\caption{Sketch of periodic dsDNA, where each bead represents one repeat 
unit consisting of one or several bases. 
(a) Many microscopic slippage events can lead to macroscopic sliding. 
(b) An applied shear force.}
\label{fig1}
\end{figure}

The dynamics of basepairing in DNA and RNA molecules plays an important 
role in biological processes such as DNA replication, transcription and 
RNA folding \cite{Alberts}. 
These dynamics can be probed in detail with modern single molecule 
techniques to exert and measure piconewton forces with nanometer 
spatial resolution \cite{SingleMolReviews}. 
For instance, double-stranded DNA (dsDNA) can be forced to open either 
by pulling on the two strands from the same end of the dsDNA (`unzipping') 
\cite{Bockelmann98,Krautbauer03,Danilowicz03} or from opposite ends 
(`shearing') \cite{Strunz99}. 
In the case of unzipping, the dynamics involves the consecutive opening 
of {\em native} basepairs, i.e. those present in the ground state of the 
molecule, and is well understood theoretically \cite{Lubensky02}. 
Here, we consider instead the shearing of dsDNA and focus specifically on 
{\em periodic} DNA sequences. 
This case is particularly interesting both from a physical and a biological 
point of view, since (i) periodic sequences have many non-native basepairing 
conformations where one strand is shifted with respect to the other, 
(ii) shearing probes the transitions between such states, i.e. the 
dynamics of DNA {\em slippage}, see Fig.~\ref{fig1}, and (iii) DNA slippage 
during genome replication allows the expansion of nucleotide repeats, 
and, for certain repeats inside genes, triggers a variety of diseases 
including Huntington's disease \cite{repetitiveDNA}. 

The mechanism for DNA slippage has already been suggested by 
P\"orschke \cite{Poerschke74}, see Fig.~\ref{fig1}a: small bulge loops 
can form at one end of the molecule when a few bases spontaneously 
unbind and rebind shifted by one or several repeat units.
Once formed, a bulge loop may diffuse along the molecule and anneal at the 
other end, effectively sliding the two strands against each other by a 
length equal to the size of the bulge loop. 
This mechanism involves only small energetic barriers compared to 
the large barrier for complete unbinding and reassociation. 
Here, we present a detailed theoretical study of 
{\em force-induced} DNA slippage, which has so far not been studied 
experimentally. 
We show that this system displays a rich dynamical behavior which can be 
controlled experimentally by adjusting the force, sequence length, and 
sequence composition. 
 
{\it Model.---} 
We consider a dsDNA of two perfectly complementary periodic 
sequences with $N$ repeat units, each consisting of $\nr$ nucleotides 
(for simplicity, we refer to repeat units also as `bases'). 
Assuming that basepairing within a strand is negligible, a basepairing 
configuration is specified by the set of the 
$n\le N$ inter-strand basepairs $\cS=\{(u_i,l_i)\}$ with 
$1\le u_1 < u_2 < \ldots < u_n\le N$ for the `upper' strand and 
analogously for the $l_i$ in the `lower' strand. 
We assign a binding energy $-\eps<0$ to each basepair and a loop cost 
$\El(j)>0$ when there are $j>0$ unpaired bases (total on both 
strands) between two consecutive basepairs. 
With a constant shear force $f$, see Fig.~\ref{fig1}b, the energy 
of a configuration $\cS$ is  
\begin{equation}
  \label{energy}
  E[\cS] = -\eps\,n[\cS] + \sum_{i=2}^{n[\cS]}\El(\Delta u_i+
            \Delta l_i-2) - f\,L[\cS] \;,
\end{equation} 
where $\Delta u_i=u_i-u_{i-1}$ and $\Delta l_i=l_i-l_{i-1}$. 
The loop cost $\El(j)$ increases with the loop length, starting from 
$\El(0)=0$. Free DNA ($f\!=\!0$) is described by  
$\El(j)=\Eini+3\,\nu\,k_BT\ln(j)$, with a loop initiation cost $\Eini>0$ 
and a logarithmic asymptotic behavior derived from polymer theory 
($\nu\approx 0.6$ is the Flory exponent) \cite{Hwa03}. 
An applied force can affect $\El(j)$, however our qualitative results are 
insensitive to its precise form \cite{later}. 
Unless stated otherwise, we keep only the constant term, 
$\El(j\!>\!0)=\Eini$, for simplicity. The total extension $L$ is 
\begin{equation}
  \label{length}
  L[\cS] = \lss (u_1\!-\!1+N\!-\!l_n) + 
           \lds\sum_{i=2}^n \min(\Delta u_i, \Delta l_i) \;,
\end{equation}
where $\lds$ and $\lss>\lds$ are the effective lengths (in the direction 
of the force) per single and double stranded unit, respectively. 
The entropic elasticity of DNA \cite{Smith96} causes both $\lds$ and 
$\lss$ to depend on the applied force, however since the DNA is almost 
fully stretched at the forces of interest here, we use the constant 
values $\lds/\nr=3.4\,\text{\AA}$ and $\lss/\nr=7\,\text{\AA}$ 
for simplicity \cite{SDNA}. 

We study the dynamics of our model both with analytical methods (described 
below) and a Monte Carlo approach using three elementary 
moves \cite{Flamm00}: opening, closing, and slippage of a 
single basepair, i.e. a pair $(u_i,l_i)$ is removed from the set $\cS$ 
or added to it, or, if the basepair is adjacent to a loop, either $u_i$ or 
$l_i$ can be changed to another base inside the loop. 
The absolute timescale of these dynamics is hard to predict, but 
comparison with bulk reannealing experiments \cite{Poerschke74} 
suggests that our simulation time step is on the order of $\mu$s in 
real time. 

\begin{figure}[tb]
\includegraphics[width=8cm]{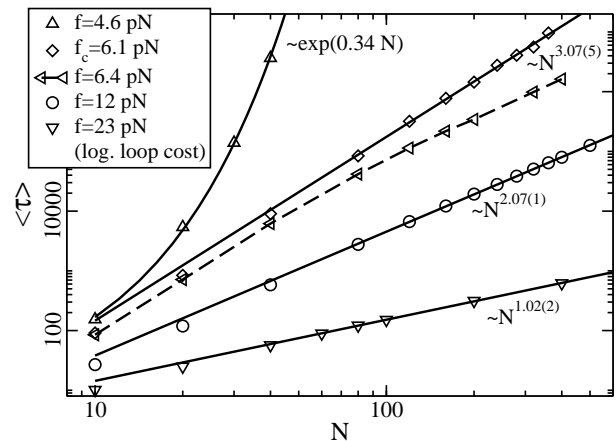}
\vspace*{-0.4cm}
\caption{\label{fig:scaling} 
Scaling of the mean rupture time $\langle\Tr\rangle$ with the number of 
bases $N$ for different shear forces (with $\eps\!=\!1.11$, $\Eini\!=\!2.8$, 
which roughly corresponds to AT-sequences at $50^\circ C$, see 
Fig.~\ref{fig:parameters}). 
The symbols represent Monte Carlo data (error less than symbol size). 
The solid lines for $f\ge \fc$ are power law fits (exponent with error in 
least significant digit is given; data with $N\le 40$ shows significant 
finite size deviations and is excluded). 
For $f<\fc$ the rupture time increases exponentially. 
The data for $f=6.4$~pN~$\gtrsim\fc$ (connected by the dashed line), 
demonstrates the crossover from diffusive to drift behavior, see main text . 
The data for $f=23$~pN is calculated including the logarithmic loop cost, 
which becomes relevant at large forces \cite{later}. 
}
\end{figure}

{\it Scaling of mean rupture times.---}
With a constant applied force $f\!>\!0$, eventually every finite dsDNA 
will rupture, since complete separation of the strands ($L\!\to\!\infty$) 
is the state of minimal free energy. 
However, both the timescale and the nature of the rupture dynamics depend 
drastically on the force.  
Fig.~\ref{fig:scaling} displays the scaling of the mean rupture time 
$\langle\Tr\rangle$ with the number of bases $N$ for a number of different 
forces (see caption for parameters). 
We observe four distinct asymptotic behaviors: an exponential increase 
with $N$ for small forces, a cubic scaling with $N$ at a certain 
force $\fc$, a nearly quadratic scaling above $\fc$ but below a second 
threshold $\fd$, and linear scaling above $\fd$. 
The behavior in the two extremes is easily interpreted: for small $f$, 
rupture is driven by thermal fluctuations across a large free energy 
barrier with an associated Kramers time that scales exponentially 
with $N$, and linear scaling at large $f$ is expected when 
individual bonds break sequentially at a constant rate.   
We now characterize the rich behavior in the intermediate 
force regime, including the nature of the two transitions. 

The thermodynamic energy barrier disappears at a force $\fc$ which 
can be estimated by balancing the binding energy per basepair with the 
mechanical work exerted when sliding both strands against each other by 
one step, 
\begin{equation}
  \label{eq:fc_naive}
  \fc \approx \eps/(2\lss-\lds) \;.
\end{equation}
$\fc$ is a critical force in the thermodynamic 
sense, if the state of complete rupture is excluded 
(see below for the exact calculation including all basepairing 
configurations). 
At $f=\fc$, the rupture dynamics is best understood by analogy with the 
reptation problem \cite{deGennes71}, since bulge loops in the DNA structure 
behave similarly to the ``stored length'' excitations of a single chain in 
a polymer network: these excitations are generated at the ends of the polymer 
with constant rate independent of $N$, diffuse along the polymer and reach 
the other end with a probability $\sim N^{-1}$. 
Therefore, the macroscopic diffusion constant for the relative motion of the 
two DNA strands should scale as $D\sim N^{-1}$ and the time for diffusion 
over distance $N$ is $\sim N^3$. 

\begin{figure}[t]
\includegraphics[width=8cm]{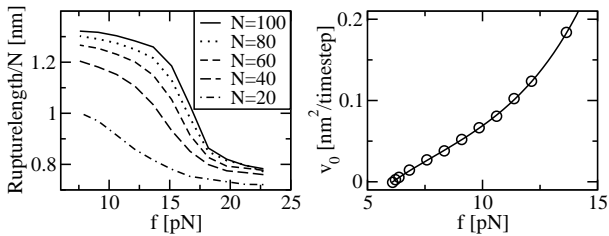}
\vspace*{-0.4cm}
\caption{(a) Rupture length as a function of applied force $f$ (parameters 
as in Fig.~\ref{fig:scaling}). 
(b) Drift coeff. $v_0(f)$ extracted from simulations with $N=150$ 
(circles) and analytical curve (solid line, $k_0=1.87$), see main text.}
\label{fig3}
\end{figure}

For $f>\fc$ strand separation is energetically a downhill process, which 
induces a drift velocity $v$ between the two strands. 
In linear response, we expext $v=\mu\,\Delta f$ for small $\Delta f=f-\fc$ 
with a mobility mediated by bulge loop diffusion, $\mu=D/k_BT\sim N^{-1}$ 
(from the Einstein relation), leading to $\langle\Tr\rangle\sim N^2$. 
Why does this behavior not persist for large forces? 
The second transition in the scaling behavior is due to a change in 
the rupture {\em mode}: 
at forces larger than $\fd \approx \eps/(\lss-\lds)$ the double 
strand can open by {\em unraveling} from both ends, i.e. the energy cost 
$\eps$ of opening a basepair at the end is outweighed by the gain 
$f(\lss-\lds)$ from a longer base-to-base distance in the single strand. 
In this unraveling mode, the rupture time scales linearly with $N$. 
The dynamical transition from sliding to unraveling is clearly 
reflected in the length at rupture, $L[\cS(\tau)]$, see 
Fig.~\ref{fig3}a, which is roughly a factor of two larger for sliding. 

\begin{figure}[tb]
\includegraphics[width=8cm]{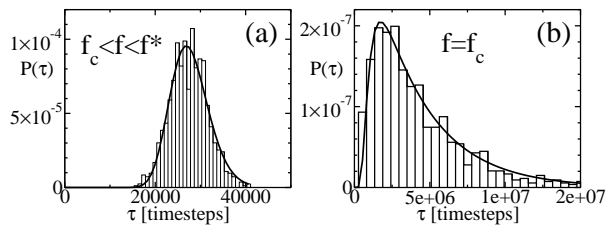}
\vspace*{-0.4cm}
\caption{\label{fig4}
Histogram of rupture times for two different forces, (a) 34.4~pN and 
(b) $\fc$=16.6~pN, but the same set of DNA parameters, $N\!=\!80$, 
$\eps\!=\!3.75$, $\Eini\!=\!2.6$, which roughly correspond to a CG-sequence 
at room temperature, see Fig.~\ref{fig:parameters}.}
\end{figure}

{\it Rupture time distributions.---}
Single molecule setups are ideally suited to record the full distribution 
of rupture times, $P(\Tr)$, which is a sensitive characteristic of the 
dynamics and permits a close examination of the physical picture introduced 
above. 
The histograms in Fig.~\ref{fig4} show $P(\Tr)$ from 
simulations at $f\!=\!\fc$ and a larger force $\fc<f<\fd$, see caption for 
parameters. 
We observe that fluctuations play a dominant role at $f\!=\!\fc$, i.e. the 
width of $P(\Tr)$ is comparable to the mean, while the rupture dynamics is 
drift dominated at the larger force, with a localized peak in $P(\Tr)$. 

To formulate the drift-diffusion dynamics quantitatively, we treat the 
number of bases in the double-stranded region as a continuum variable $x$ 
with $0<x<N$, and consider the probability distribution 
$\cP(x,t)$, which satisfies the continuity eq. 
$\partial_t \cP(x,t) = -\partial_x j(x,t)$ with a force-dependent current  
\begin{equation}
  \label{macro_current}
  j(x,t) = -D(f,x)\,\partial_x \cP(x,t) - v(f,x)\,\cP(x,t)\;.
\end{equation} 
The above discussion suggests a diffusion coefficient of the form 
$D(f,x)=D_0(f)/x$ and similarly a drift $v(f,x)=v_0(f)/x$. 
We have an absorbing boundary at $x=0$ and it is natural to choose a 
reflecting boundary at $x=N$ and a delta peak at $x=N$ as initial condition. 
The solution $\cP(x,t)$, which must in general be obtained numerically, 
determines the rupture time distribution through $P(\Tr)=j(0,\Tr)$. 

We can determine the force-dependence of the diffusion coefficient and drift 
empirically by fitting the calculated $P(\Tr)$ to the simulation data using 
$D_0$ and $v_0$ as adjustable parameters. 
The solid lines in Fig.~\ref{fig4} show that the 
drift-diffusion theory describes the simulation data well. 
Fig.~\ref{fig3}b shows the fitted $v_0$ as a function of $f$ (circles). 
The drift vanishes at the critical force, $v_0(\fc)=0$, 
confirming the physical picture. 
The drift-diffusion theory also explains the crossover behavior in the 
vicinity of $f\!=\!\fc$, see Fig.~\ref{fig:scaling}: the drift is 
significant only when the system size $N$ is larger than the diffusive 
length $D_0/v_0$ \cite{Lubensky99}. 
Hence, with $v_0\sim\Delta f$, reptation-like dynamics is expected in a 
force interval $\delta f\sim N^{-1}$ around $\fc$.

{\it Microscopic dynamics.---}
Next, we study how the macroscopic drift in eq.~(\ref{macro_current}) 
emerges from the microscopic bulge loop dynamics and determine $v_0(f)$ 
in terms of our system parameters. 
Since bulge loops on opposite strands annihilate each other when they 
meet, the bulge loop dynamics is equivalent to a reaction-diffusion 
system of particles and antiparticles in one dimension. 
Both particles and antiparticles are created at each end, however with 
different rates determined by the applied force. 
{From} the underlying master equation for these processes one obtains the 
mean-field equations \cite{later} 
\begin{eqnarray}
  {\partial_t}u(y,t) &=& k_0 {\partial^2_y}u(y,t)-k_1 u(y,t)\,l(y,t)+k_2 \;,
  \nonumber \\ \label{eq:meanfield}
 {\partial_t}l(y,t) &=& k_0 {\partial^2_y}l(y,t)-k_1 u(y,t)\,l(y,t)+k_2 \;,
\end{eqnarray}
where $u(y,t)$ and $l(y,t)$ denote the bulge loop density on the 
upper/lower strand, $y\in [0,x]$ is the position 
within the double stranded region, and $k_0$, $k_1$, $k_2$ are the 
rates for hopping, annihilation, and pair creation, respectively. 
At the boundaries, the densities take on constant 
values, $u(0,t)=l(x,t)=\rho_<$ and $u(x,t)=l(0,t)=\rho_>$, where 
$\rho_<(f)$ and $\rho_>(f)$ are calculated below by assuming a local 
equilibrium of the DNA at the edges. 
The macroscopic drift is determined by the stationary solution and 
depends only on the difference between the loop densities on the 
upper/lower strand, $v(f,x)=k_0{\partial_y}[u(y)-l(y)]$. 
Using eq.~(\ref{eq:meanfield}) this yields 
$v_0(f)=2k_0[\rho_>(f)-\rho_<(f)]$. 
Fig.~\ref{fig3}b shows that this result is in excellent agreement with the 
empirical $v_0(f)$ obtained above. 

Since the loop cost $\El(j)$ is larger for two separate loops than for a 
single one of the combined length, bulge loops on the same strand feel a 
short-range attraction. However, the interaction is not strong enough to 
cause a significant aggregation of the loops in our Monte Carlo simulations. 
This is consistent with the observation that with our DNA parameters, the 
interaction energy $\Eini$ is never significantly larger than the entropic 
cost $\sim\log\rho$ of colocalization at loop density $\rho$. 
While $v_0(f)$ is apparently robust to interaction effects, the diffusion 
coefficient $D_0(f)$ is sensitive to interactions as well as correlations. 
Both are neglected in eq.~(\ref{eq:meanfield}), leaving  
the microscopic calculation of $D_0(f)$ as a challenge for the future.



\begin{figure}[t]
  \includegraphics[width=\columnwidth]{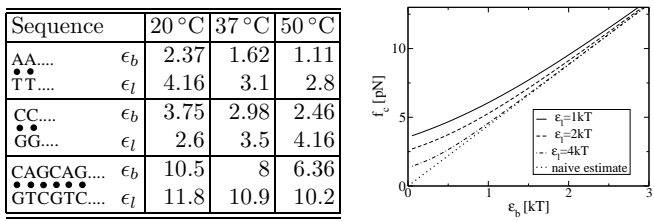}
\vspace{-0.5cm}
\caption{\label{fig:parameters}
(a) Model parameters for different DNA sequences and temperatures as obtained 
by fitting to a detailed thermodynamic model \cite{Zuker03,later}  
(all energies in units of $kT$). 
(b) The exact critical force compared to the estimate of 
eq.~(\ref{eq:fc_naive}).}
\end{figure}

{\it Critical force.---}
To obtain the exact critical force, we need the partition function 
$Z=\sum e^{-E[\cS]/k_BT}$ summed over all configurations $\cS$ with 
at least one basepair. It is useful to allow for different numbers of 
bases in the two strands, e.g. $1\le u_i \le N$ and 
$1\le l_i \le M$, with a corresponding partition function 
\begin{equation}
  \label{eq:Z}
  Z(N,M)=\sum_{i=0}^{N-1}\sum_{j=0}^{M-1}\bss^{i+j}
         \sum_{n=1}^{N-i}\sum_{m=1}^{M-j} Z_p(n,m) \;,
\end{equation}
where $\bss=e^{f\lss/k_BT}$ is the Boltzmann factor for a stretched 
base, and $Z_p(n,m)$ is the partition function for the central, 
double stranded section starting with the first and ending with the 
last basepair, cf. Fig.~\ref{fig1}b. 
We calculate $Z_p(n,m)$ recursively by introducing a complementary  
partition function $Z_u(n,m)$ containing only structures where the 
last of the $n$ upper bases is {\em not} bound to the last of the $m$ 
lower bases: 
\begin{eqnarray}
  \label{eq:recursion}
  Z_p(n\!+\!1,m\!+\!1) &=& q\,\bds\, Z_p(n,m) + q\,\bds\,g\, Z_u(n,m)\;, \\
  Z_u(n\!+\!1,m\!+\!1) &=& g\sum_{k=1}^n Z_p(k,m\!+\!1) +
               g\sum_{k=1}^m Z_p(n\!+\!1,k) \nonumber \\
  & & +\; g\,\bds Z_p(n,m) + \bds\,Z_u(n,m) \;. \nonumber 
\end{eqnarray}
Here, the Boltzmann factors $q=e^{\eps/k_BT}$, $g=e^{-\Eini/2k_BT}$, 
and $\bds=e^{f\lds/k_BT}$ account for basepairing, loop costs, and 
stretching of double strand, respectively. 
These recursion relations allow the efficient 
calculation of the partition function for finite $N,M$. To obtain the 
critical behavior for $N\to\infty$, we take the 
z-transform $\Zh(z,y)=\sum_{N,M}\,Z(N,M) z^N y^M$. 
The inverse z-transform is then determined by the simultaneous poles 
of $\Zh(z,y)$ in $z$ and $y$, and for large $N$ the pair of poles 
with the smallest $|zy|$ dominates. 
A detailed analysis of the critical behavior will be presented 
elsewhere \cite{later}; here we are interested in $\fc$, i.e. the force 
where the dominant pole switches. We find that $\fc$ 
is the nontrivial root of 
\begin{equation}
  \label{eq:fc_exact}
  \left(\frac{\bss^2}{\bds}-q\right)\left(\frac{\bss^2}{\bds}-1\right)-
  g^2q\left(\frac{2}{\bss-1}\frac{\bss^2}{\bds}+1\right)=0 \;. 
\end{equation}
When $\eps$ or $\Eini\gg k_BT$, the second 
term is negligible and the nontrivial root of (\ref{eq:fc_exact}) is 
$\bss^2/\bds=q$, recovering the naive estimate (\ref{eq:fc_naive}). 
However, for smaller $\eps$, $\Eini$ one finds significant deviations from 
(\ref{eq:fc_naive}), see Fig.~\ref{fig:parameters}. 


{\it Loop densities.---}
Using the same approach as for the calculation of $\fc$, we can calculate 
the loop densities $\rho_<$, $\rho_>$ introduced above. 
Assuming local equilibration at the edges, i.e. equilibration between all 
possible conformations of the two strands with a fixed central basepair, we 
find $\rho_< = \sum_{a,b} P(a,b)\,a/\lambda$ and 
$\rho_> = \sum_{a,b} P(a,b)\,b/\lambda$, where $\lambda={\min(a,b)+1}$ and 
$P(a,b)={\bss}^{b-a}{\bds}^\lambda q\,g\,
Z_p(N\!-\!b\!-\!1,N\!-\!b\!-\!1)/Z_p(N,N)$. 
The sums can be evaluated exactly for large $N$ \cite{later}.


{\it Conclusions.---}
We find a response of periodic dsDNA to shear forces that is very distinct 
from that for nonperiodic sequences. 
Above a thermodynamic critical force $\fc$, but below a {\em dynamic} 
critical force $f^*$, bulge loop diffusion allows periodic DNA to open by 
{\em sliding}. 
This mechanism leads to a much lower thermodynamic critical force than the 
{\em unraveling} mechanism by which nonperiodic DNA opens. 
Within our model, we have calculated $\fc$ exactly and characterized the 
associated dynamics, which is effectively {\em viscoelastic} with a creep 
compliance $\sim N^{-1}$ for $\fc<f<f^*$. 
Above $f^*$ periodic dsDNA also opens predominantly by unraveling 
(this dynamical transition may be regarded as a remnant of the 
thermodynamic transition for nonperiodic sequences). 
Interestingly, periodic DNA could be used as a viscoelastic 
nanomechanical element with properties that are {\em programmable} by 
choosing sequence length and composition.
This may lead to applications in microstructured devices, similar to the 
programmable DNA-based force sensors reported in Ref.~\cite{Albrecht03}.  


\begin{acknowledgments}
We thank T. Hwa, F. K\"uhner, and M. Rief for fruitful discussions and 
the {\it DFG} for financial support. 
\end{acknowledgments}

\end{document}